# Long Lived NMR Signal in Bone

Boyang Zhang[1], Jae-Seung Lee[1,2], Anatoly Khitrin,[3] and Alexej Jerschow[1*]

[1]Chemistry Department, New York University, New York, NY 10003

[2]Center for Biomedical Imaging, Radiology Department, New York University, New York, NY 10003

[3]Department of Chemistry, Kent State University, Williams Hall, Kent, OH 44242.

* Correspondence and request for materials should be addressed to:

Alexej Jerschow
Chemistry Department
New York University
New York, NY 10003
phone: 212 998 8451
alexej.jerschow@nyu.edu

**Classification**
Physical Sciences: Applied Physical Sciences, Chemistry

**Abstract**

Solids and rigid tissues such as bone, ligaments, and tendons, typically appear dark in magnetic resonance imaging (MRI), which is due to the extremely short-lived proton nuclear magnetic resonance (NMR) signals. This short lifetime is due to strong dipolar interactions between immobilized proton spins, which render it challenging to detect these signals with sufficient resolution and sensitivity. Here we show the possibility of exciting long-lived signals in cortical bone tissue with a signature consistent with that of bound water signals. Contrary to long-standing belief, it is further shown that dipolar coupling networks are an integral requirement for the excitation of these long-lived signals. The use of these signals could enhance the ability to visualize rigid tissues and solid samples with high sensitivity, resolution, and specificity via MRI.

Magnetic resonance imaging (MRI) is a prime noninvasive diagnostic tool in the medical field (1, 2), and is becoming popular in the materials sciences as well (3-7). One of its major challenges, in particular in the medical field, is the requirement of relatively long-lived signals (> 1-10 ms) for adequate image reconstruction with sufficient microscopic detail *in vivo*. Bone and rigid tissues, such as ligaments and tendons, however, have very short-lived signals (< 1ms), and therefore often remain dark in conventional MRI images (8-12). Although there exist advanced methods for visualizing relatively short-lived signal components, these are generally demanding on hardware, require specialized instrumentation, and are frequently incompatible with *in vivo* constraints (8-10).

The origin of the short lifetime in rigid tissues (and rigid samples, in general), can be two-fold. First, considerable magnetic field inhomogeneity can be present due to susceptibility gradients, leading to signal dephasing (inhomogeneous broadening) (13). The second contribution is from homogeneous broadening, with one of its major mechanisms being the dipolar coupling between many spins. The inhomogeneous mechanism can, in principle, be refocused by echo pulse sequences. Elaborate schemes, typically based on interleaved pulsing and acquisition, often in combination with sample spinning, have been developed for reducing the homogeneous broadening (14). These techniques are not practical for many applications, especially not for *in vivo* MRI (15, 16).

One can excite narrow portions of an inhomogeneously broadened resonance by selective radiofrequency (rf) pulses, while it had been a long-standing dogma that the same cannot be done for dipolar broadening (17) – until it was shown to be possible by Khitrin *et al*. (18-20). In this work, it was shown that one could excite many narrow signals (up to a thousand), from a dipolar broadened line in the context of 'molecular photography' (21, 22). Since then, it was shown that narrow (and therefore long-lived) signals could be excited in a number of solid compounds, including adamantane, 5-CB liquid crystal, naphthalene, polybutadiene, and glucose, by either a long and weak radiofrequency (rf) irradiation, or such irradiation in combination with a strong refocusing $\pi$ pulse (19, 20). A complete theory of this puzzling phenomenon is still missing. The proposed mechanism for the appearance of these long-lived signals is based on a symmetry breaking which comes from a small difference of chemical shifts of nuclei directly coupled by much stronger dipole-dipole interactions (19, 20).

In this Letter, we show that such long-lived signals can be excited in cortical bone tissue samples and used for imaging. Most notably, it is shown here that the intensity of the long-lived signals is related to the dipolar coupling network complexity, further underlining that this new NMR phenomenon is caused by unusual spin dynamic properties. Fig. 1 shows respective experiments on a cortical tissue sample. Here, two sequences are compared, one with a long and weak rf irradiation, and one with a shorter irradiation followed by a hard $\pi$ pulse which is phase-shifted by $\pi/2$. We call the first one the long-lived response (LLR) signal, and the second one the LLR echo (LLRE) signal. The excited signal generally appears at the same frequency as the respective weak rf irradiation, and the linewidths are reduced when compared to the single hard pulse excitation by factors of 16 and 7, respectively (from 1150 Hz for a hard pulse spectrum). The LLRE signal is stronger than the LLR signal by a factor of 7. Both are significantly weaker than the hard pulse spectrum. The decreased sensitivity is, however, more than compensated for the flexibility provided by longer signal lifetimes in MRI practice. Another important point is also that the LLRE signal can be clearly linked to the bound water pool (see below), and does not respond strongly to the pore water signal which would also be included in the amplitude of the conventional signal. Further examples of LLRE signal excitation in a homogeneously-broadened system are shown in the supplementary information section, including a demonstration with a solid collagen sample (supplementary Fig. S1).

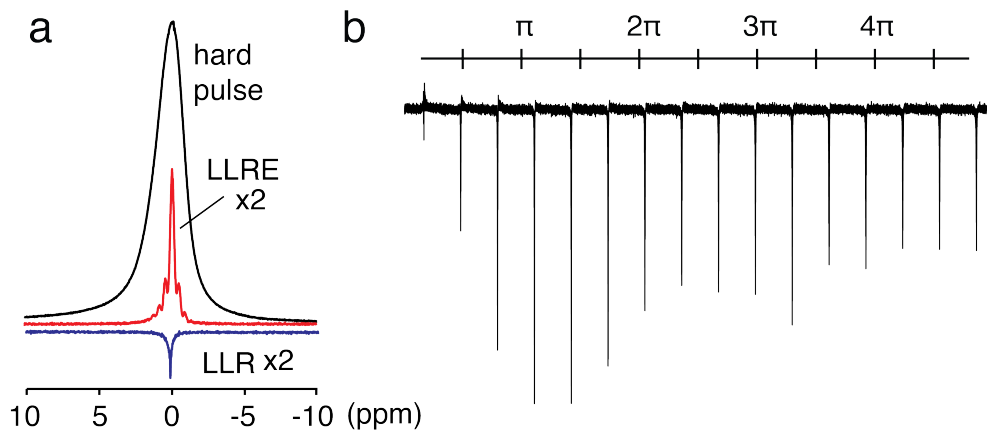

**Fig. 1.** Long-lived signals excited in a cortical bone sample. (**a**) Conventional $^1$H NMR spectrum (black) compared with a LLR spectrum (blue) and a LLRE spectrum (red) of a cortical bone sample. The linewidths are 1150 Hz, 70 Hz, and 165 Hz, respectively. The LLR spectrum was acquired with after applying a 18 ms long pulse with rf power of 40 Hz. The LLRE spectrum was acquired with a 5 ms soft pulse with rf power of 80 Hz, followed by a high power $\pi$ pulse. (**b**) LLR signals excited with pulses of 40 Hz power with nominal flip angles ranging from ~$\pi/6$ to $5\pi$ in a cortical bone sample.

As a testament to the homogeneous character of the origin of the LLR signal, a rf nutation curve is shown in Fig. 1b. Under conditions of inhomogeneous broadening, the signal on-resonance with the pulse would be modulated by the rf nutation frequency. Under homogeneous broadening, however, even if the signal were on-resonance with the rf irradiation, no oscillation should be seen beyond a regime where the average dipolar coupling strength $\langle H_D / \hbar \rangle$ exceeds the bandwidth of the rf irradiation. This is indeed observed in this experiment.

Multi-spin quantum simulations are particularly challenging to perform for systems larger than 10-13 densely coupled spins (23). A limited simulation with 10 dipolar-coupled spins, however, also shows the appearance of a LLR signal (Fig. 2a). The rf nutation curve of Fig. 2b also reveals a behavior similar to the experimental results of Fig. 1b, thus illustrating the dipolar coupling nature of the observed effects.

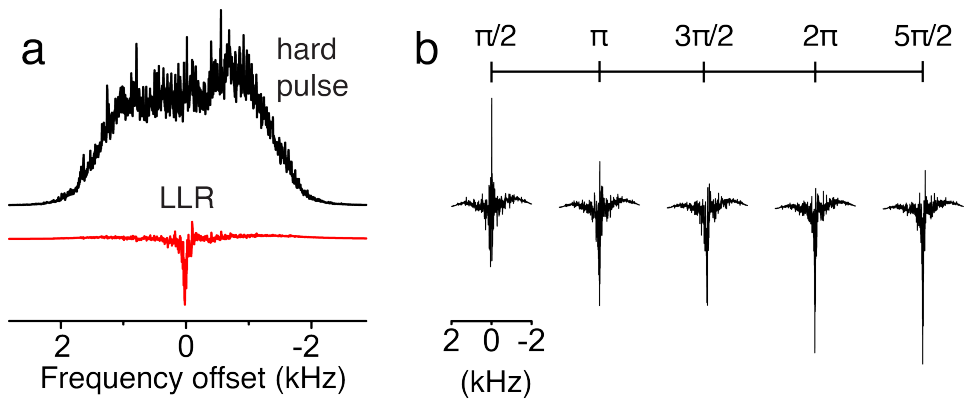

**Fig. 2**. Simulated spectra of long-lived signal excitation in 10 dipolar-coupled spins. (**a**) Comparison between a conventional spectrum with a π/2 excitation pulse of 25 kHz power (black) and a LLR spectrum (red) using a nominal 2π pulse with 50 Hz power. (**b**) Simulated LLR spectra vs. the flip angle of the excitation pulse (at 50 Hz pulse power).

In order to provide further insight into the origin of the LLR phenomenon, the following experiment was performed: a cortical bone sample was equilibrated successively in $D_2O/H_2O$ mixtures for a prolonged period (24h) at 55ºC in order to facilitate the equilibration of the solution with pore and bound water pools in the tissue. This procedure can be used to determine the connection between the short-lived NMR signals and bound bone water fractions (24, 25). In previous work, it was shown that this signal, when measured *in vivo*, correlated with cortical bone porosity (24, 25). In Fig. 3 it is seen that the conventional pulsed NMR signal obeys a linear relationship with $D_2O$ concentration, while the long-lived signal displays a strikingly non-linear relationship. The long-lived signal decreases rapidly when the $D_2O$ content increases over the first 10%, and then decreases more gradually when increasing the $D_2O$ content further.

Conventional NMR theory does not provide any clue as to why there may be such a rapid decrease of the LLR signal in the low $D_2O$ percentage range. These findings are therefore a strong indication for an interpretation based on the dipolar coupling network: The LLR intensity decreases with the decreasing dipolar network connectivity due to dilution with $D_2O$ and H/D exchange. The exchange dynamics itself may play a further role in destroying the averaged coupling network, and hence also the LLR signal.

This result presents a new, and perhaps the strongest demonstration so far, that the long-lived signals depend on dipolar coupling complexity. With respect to the previously-reported distinction between cortical bone pore water and bound water (26), this experiment also clearly demonstrates that the LLR signals must arise from bound water (in pore water, the major broadening arises from the inhomogeneous mechanism, which is clearly ruled out in the results of Figs. 1 and 3).

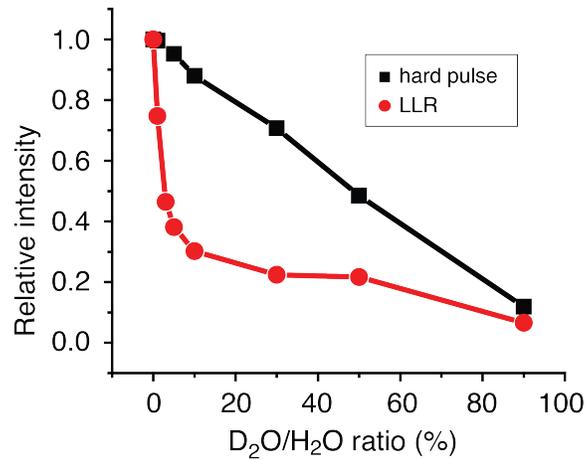

**Fig. 3.** Signal intensity as a function of $D_2O/H_2O$ ratio for a hard pulse response (black squares) and LLR (red circles) in a cortical bone sample. The rapid decrease of the LLR signal with $D_2O$ ratio increase is a testament to the dipolar coupling nature of the effect.

The line narrowing afforded by this LLR signal excitation can then be used for imaging with enhanced resolution, as illustrated in Fig. 4. Here, it is seen that the LLRE image (Fig. 4a) of cortical bone appears superior to the conventional gradient echo (GE) image of the same nominal resolution (Fig. 4b). Furthermore, it is seen that in the GE the image is distorted, likely as a result of the broad linewidth. The reduced linewidth in LLRE places fewer constraints on the minimal gradient strength necessary to achieve a specified resolution. Image readout times can also be made longer, thereby obviating the need for strong and fast-switching gradients. These long-lived signal acquisitions could be combined with radial (27) or other nonuniform (28) readout techniques, where they could also be used to reduce the required gradient strength along the individual *k*-space trajectories. The one-dimensional imaging projections illustrate this point further (Fig. 4c). While the conventional spin-echo (SE) image shows the strongest signal, its resolution is significantly degraded due to signal overlap. In particular, both the GE and the SE show signal extending beyond the physical limits of the sample on the right side. While it appears here that the SE shows the best signal, it should be remembered that the signal is only stronger at the expense of resolution, and that such efficient echoes cannot normally be implemented on an MRI scanner. Additional demonstration of resolution enhancement by LLRE is demonstrated in supplementary Fig. S2 for a trabecular bone sample. Supplementary Fig. S3 also shows how LLRE allows one to observe cortical bone signals in a sample containing both cortical and trabecular bone.

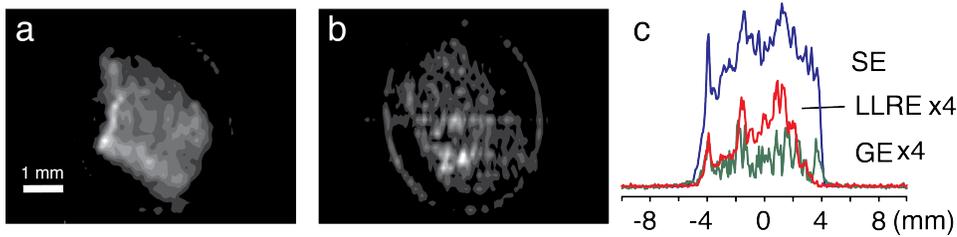

**Fig. 4.** Images of a cortical bone sample. (**a**) Two-dimensional LLRE and (**b**) conventional gradient echo (GE) images, both at nominal resolutions of 51×151 μm$^2$, with 51 μm as the resolution of the direct imaging dimension. (**c**) One-dimensional images using a spin-echo (SE), gradient echo (GE), and a LLRE. The SE image is scaled down by a factor of 4 for comparison.

In conclusion, we show herein that long-lived signals can be excited in the homogeneously broadened signal portion of cortical bone. This signal originates therefore from bound water (as opposed to pore water), due to the homogeneous nature of the spectrum. Furthermore, deuterium-exchange experiments indicate that the long-lived signals correlate with the dipolar coupling network complexity, which has never been seen directly before. These long-lived signals could become very important for imaging rigid tissues and bone, in particular, *in vivo*, where they may become useful for the diagnosis and the study of osteoporosis and many other bone disorders.

**Methods**

Samples of fresh bovine tibia bone were purchased from a local slaughterhouse. Cortical bone was obtained from the center of the tibia and soaked in PBS solution for three hours before experiments. Trabecular bone was cut from the end of the tibia using a 3-inch bone cutter (Apiary Medical, San Diego, CA). The bone samples were cut to fit into a 5 mm NMR tube. Dry collagen strips (from Achilles tendon CAS 9007-34-5) and 99.8% D$_2$O were purchased from Sigma-Aldrich (St. Louis, MO, USA) and used without further modification. All experiments were performed at 11.7 T (500 MHz $^1$H frequency) using a Bruker Avance spectrometer (Billerica, MA, USA) equipped with a BBO probe. To excite the LLR signal, a long and weak pulse was applied. The duration and power were optimized for maximal signal and were 18 ms and at a rf amplitude ($\gamma B_1/2\pi$) of 40 Hz. A pre-acquisition delay of 1 ms was used, which improved the baseline in the spectra by dephasing spurious broader components. The LLRE signal was excited first with a 5 ms soft pulse with rf power of 80 Hz, followed by a π pulse with a 25 kHz power. The nutation experiments for LLR in Figs. 1B and S1b were performed with flip angles ranging from ~π/6 to 5π with a pulse power of 40 Hz.

For the simulation of Fig. 2, a system of 10 dipolar-coupled spins ½ with random dipolar couplings (in the range spanning -750 Hz ~ +900 Hz) and random chemical shifts (spanning -400 Hz ~ +460 Hz) was chosen to mimic a homogeneously broadened system with a continuous spectrum, while avoiding any accidental symmetries. After a rectangular pulse was applied to the thermal equilibrium state, a 2048-point free induction decay was numerically calculated, exponential multiplication with 10 Hz applied, and the corresponding spectrum was obtained through Fourier transform.

The equilibration of bone samples in $D_2O/H_2O$ mixtures (Fig. 3) was conducted following the procedure reported by Techawiboonwong et al. (24). The bone sample was immersed in $D_2O/H_2O$ mixtures of varying isotopic composition (1%, 3%, 5%, 7%, 10%, 30%, 50% and 90% $D_2O$ volume fraction) for 24 hours at 55°C. To remove the free water in large macroscopic pores on the endosteal surface, the specimen was blow-dried for 1 minute. The sample was then placed inside a 5-mm NMR tube. Fluorinated oil (Fluorinert, FC-77, 3M, St. Paul, MN, USA) was filled into the void spaces for protection and for reduction of susceptibility artifacts.

The two-dimensional LLRE images (Figs. 3a, 3b and S3) were acquired with a gradient echo imaging block with a 2 ms echo time and 16 averages. In the 2D images of cortical bone (Figs. 3a and 3b), the data matrix 128×32 had a resolution of 51×151 $\mu m^2$. The applied gradient strengths were 7.8 G/cm for both the read-out gradient and the phase-encoding gradient. In supplementary Fig. S3, the data matrix was of size 256×32 with a resolution of 77×226 $\mu m^2$; the applied gradient strengths were 6 G/cm for the read-out gradient and a maximum of 5.2 G/cm for the phase-encoding gradient. The one-dimensional images in Figs. 3c and S2 were obtained along the *z*-direction, i.e., along the sample tube and with 2 ms echo time and 2 averages. The gradient strength was 6 G/cm and the spatial resolution 76.5 μm/pt.

## Acknowledgements

We acknowledge funding from US National Science Foundation under Grant No. CHE0957586 (to AJ) and the National Institutes of Health under Grant No. K25AR060269 (to JSL). The experiments were performed in the Shared Instrument Facility of the Department of Chemistry, New York University, supported by the US National Science Foundation under Grant No. CHE0116222.

**Supplementary Information for**
*Long Lived NMR Signal in Bone*
by Boyang Zhang, Jae-Seung Lee, Anatoly Khitrin, and Alexej Jerschow

**Supplementary Figures**

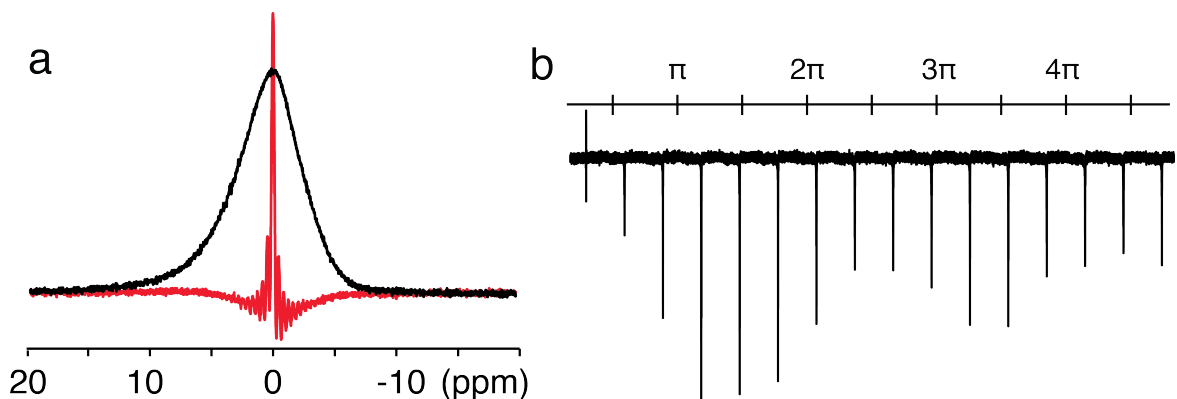

**Fig. S1.** Conventional and long-lived response $^1$H spectra of a dry collagen sample. (**a**) Hard-pulse excitation (black) and LLRE spectra with a pulse power of 40 Hz (red). (**b**) LLR signals were excited with pulses with flip angles ranging from ~π/6 to 5π with a pulse power of 40 Hz.

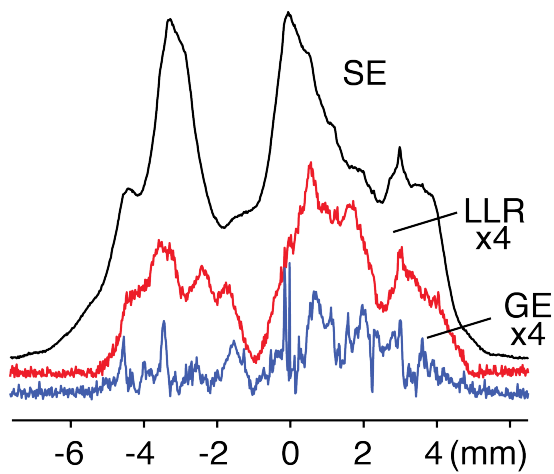

**Fig. S2.** One-dimensional images of a trabecular bone sample with conventional spin echo (SE) (black), LLR (red) and conventional gradient echo (GE) (blue).

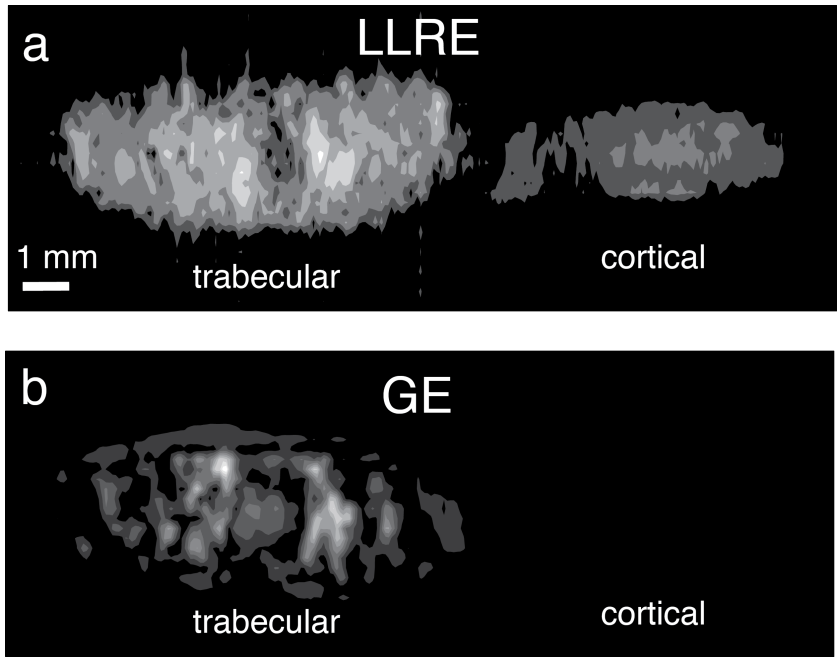

**Fig. S3.** Two-dimensional images of a sample containing a piece of cortical and trabecular bone. (**a**) LLRE and (**b**) conventional GE. Resolution: 77×226 μm$^2$.